# A nuclear-quantum-corrected machine-learning potential reveals quantum-enhanced hydrogen segregation at general grain boundaries in α-iron

Kazuma Ito*

Advanced Technology Research Laboratories, Nippon Steel Corporation, 20-1 Shintomi, Futtsu, Chiba 293-8511, Japan

* Corresponding author. Advanced Technology Research Laboratories, Nippon Steel Corporation, 20-1 Shintomi, Futtsu, Chiba 293-8511, Japan

Email: ito.nn3.kazuma@jp.nipponsteel.com (K. Ito)

**Abstract**

Atomistic descriptions of hydrogen diffusion and trapping at defects are essential for understanding hydrogen embrittlement. As the lightest solute in metals, hydrogen exhibits nuclear quantum effects that alter these processes even at room temperature. Explicit treatment of such effects is computationally demanding, limiting large-scale simulations of complex environments. Here, we use an Fe–H machine-learning interatomic potential (MLIP) based on the performant implementation of the atomic cluster expansion (PACE), covering diverse Fe–H environments, and relabel the training configurations underpinning its transferability with quantum mean forces from centroid-constrained path-integral molecular dynamics at 300 K. This yields a nuclear-quantum-corrected PACE (NQC-PACE) without additional density functional theory calculations. At parent PACE, NQC-PACE describes nuclear quantum effects on hydrogen trapping at vacancies, dislocations, surfaces and general grain boundaries, H–H interactions, and diffusion in α-Fe. Grand-canonical Monte Carlo/molecular dynamics simulations show nuclear quantum effects markedly enhance hydrogen segregation at general grain boundaries and trapping behaviour in closer agreement with experimental trends. This enhancement arises from selective quantum stabilisation of open, anisotropically soft local environments. Our framework uses finite-temperature quantum mean forces to relabel the configurational space covered by an MLIP, enabling large-scale analysis of complex materials where light-element quantum effects matter.

## Introduction

Green hydrogen is expected to serve as a key energy carrier in a decarbonised society, but ensuring the reliability of the structural materials used in its production, transportation, storage and utilisation remains a major challenge[1,2]. Increasing material strength is essential for improving hydrogen-storage density and reducing the weight of structural components, yet higher strength generally increases susceptibility to hydrogen embrittlement[3,4]. Hydrogen diffusion through materials, together with its trapping and local accumulation at lattice defects such as vacancies, dislocations and grain boundaries, plays a central role in the processes leading to embrittlement[1,2,5-8]. An atomistic understanding of hydrogen stability and dynamics across diverse defect environments is therefore essential[9,10]. Intergranular fracture is particularly important in high-strength steels, motivating the need to clarify hydrogen segregation and trapping at general grain boundaries—that is, boundaries lacking specific crystallographic symmetry[11] and widely present in practical polycrystalline materials[3,7,12].

Because hydrogen in metals is both light and highly mobile, its local concentration and dynamic behaviour during deformation are difficult to observe directly in experiments[13]. Density functional theory (DFT) calculations[3,14-17] and atomistic simulations based on interatomic potentials[18-21] have therefore been widely used to investigate hydrogen diffusion and interactions with lattice defects. Although DFT calculations offer high accuracy, their accessible system sizes and timescales are insufficient for directly treating general grain boundaries containing numerous inequivalent sites and multiple hydrogen atoms. Recent advances in machine-learning interatomic potentials (MLIPs) have enabled large-scale simulations while retaining near-DFT accuracy, leading to the development of MLIPs for a wide range of hydrogen-containing metallic materials[22-28]. In the Fe–H system, neural-network MLIPs have provided a unified description of diverse environments relevant to hydrogen embrittlement, including bulk crystals, vacancies, dislocations, surfaces, symmetric grain boundaries and general grain boundaries, thereby enabling large-scale analyses of hydrogen segregation and the deformation and fracture behaviour of general grain boundaries[28]. The same comprehensive training dataset was subsequently used to construct an MLIP based on the performant implementation of the atomic cluster expansion (PACE)[28], preserving this structural coverage and accuracy while substantially accelerating the calculations[27]. This model established a practical basis for simulations of hydrogen diffusion, segregation, deformation and fracture.

Hydrogen is, however, the lightest solute element in metals and is affected even at room temperature by nuclear quantum effects arising from zero-point motion and spatial nuclear delocalisation[29,30]. The MLIPs for hydrogen-containing metallic materials described above[22-28] generally represent the Born–Oppenheimer potential-energy surface and treat hydrogen nuclei as classical particles, thereby excluding nuclear quantum effects unless they are introduced separately. Path-integral molecular dynamics (PIMD) and related methods can evaluate finite-temperature quantum nuclear statistics, and have shown that nuclear quantum effects modify hydrogen diffusion in α-Fe and trapping at lattice defects[29,30]. In PIMD, however, each quantum nucleus is represented by multiple beads, requiring sufficient equilibration and statistical sampling[29,31]. Direct application of PIMD to large atomistic models, including those of general grain boundaries, therefore remains prohibitively expensive[30,32,33].

To reduce this computational burden, methods have been proposed that coarse-grain the ring-polymer degrees of freedom and map the quantum statistics generated by PIMD onto lower-dimensional machine-learned effective potentials[33,34]. In molecular systems, quantum centroid forces obtained from centroid-constrained PIMD, or their differences from classical forces, have been used to train effective potentials that reproduce equilibrium quantum statistics or approximate quantum dynamics at costs close to those of classical molecular dynamics[33-38]. Related single-replica mappings have been demonstrated for model potentials, molecular water, proton-transfer systems and liquid water[33,34]. However, these approaches have been examined mainly in molecular systems, and their extension to MLIPs for metals spanning broad configurational spaces remains largely unexplored. MLIPs for metals are often required to describe, within a single model, atomic environments with markedly different local coordination and strain states, including perfect crystals, vacancies, dislocations, surfaces and grain boundaries[27,28]. In particular, no established strategy exists for constructing training data that consistently incorporate finite-temperature nuclear quantum effects across the diverse atomic environments relevant to hydrogen embrittlement.

Here, we reuse an existing high-speed PACE-based MLIP[27] that describes diverse Fe–H environments relevant to hydrogen embrittlement, together with the comprehensive training dataset underpinning its broad structural coverage[28]. For each hydrogen-containing training configuration, we perform centroid-constrained PIMD at 300 K while retaining the original hydrogen position as the centroid coordinate, and relabel the same configurational space with the resulting quantum mean forces. This procedure yields a nuclear-quantum-corrected PACE (NQC-PACE). It introduces 300 K nuclear quantum corrections throughout the broad Fe–H configurational space described by the parent PACE without requiring additional DFT calculations, while retaining the same computational cost as the parent model and allowing direct application in conventional molecular dynamics and grand-canonical Monte Carlo/molecular dynamics (GCMC/MD) simulations[39]. We validate NQC-PACE for hydrogen trapping at diverse lattice defects, H–H interactions and hydrogen diffusion in α-Fe through comparisons with explicit PIMD, mass thermodynamic integration (mass TI)[40], DFT-based vibrational calculations and experiment. GCMC/MD simulations of general grain boundaries further show that nuclear quantum effects markedly enhance grain-boundary hydrogen segregation, and reveal site-selective quantum stabilisation in open, anisotropically soft local environments as its structural origin. By relabelling the configurational space covered by an existing MLIP with finite-temperature quantum mean forces, this work provides a practical framework for large-scale analyses of nuclear quantum effects across complex atomic environments involving lattice defects, interfaces and diffusion processes.

## Results

### Construction of NQC-PACE incorporating hydrogen nuclear quantum effects

Figure 1 outlines the construction of NQC-PACE, which incorporates the nuclear quantum effects of hydrogen at 300 K. The model was built on an existing PACE potential capable of describing diverse Fe–H atomic environments relevant to hydrogen embrittlement[27] (Fig. 1a). The training dataset for this PACE was constructed by using strained bulk crystals, generalised stacking faults, surfaces, symmetric tilt grain boundaries and vacancy clusters, among others, as initial structures and efficiently and comprehensively sampling atomic environments relevant to hydrogen embrittlement in α-Fe through a concurrent-learning strategy[28,41]. The resulting PACE reproduces the fundamental properties of α-Fe, lattice defects and their interactions with hydrogen with DFT-level accuracy. It has also been shown to exhibit high transferability to screw and edge dislocations, which were not explicitly included in the training data, as well as to the deformation and fracture processes of hydrogen-segregated general grain boundaries[27].

Hydrogen-containing atomic configurations were extracted from the PACE training dataset, and PIMD simulations were performed at 300 K while constraining the position of each hydrogen atom in the original configuration as the centroid of its ring polymer. This enabled the quantum delocalisation of the hydrogen nuclei to be sampled while retaining, as centroid coordinates, the local atomic environments underpinning the transferability of the parent PACE. The resulting quantum mean forces acting on the hydrogen atoms and surrounding Fe atoms were then obtained (Fig. 1b). In other words, the hydrogen-containing configurational space of the parent PACE was relabelled with quantum mean forces at 300 K. This design was intended to extend nuclear quantum corrections across the diverse atomic environments relevant to hydrogen embrittlement while preserving the broad configurational coverage of the parent training dataset.

NQC-PACE was constructed using the resulting quantum mean forces as the primary training targets and the parent PACE as the initial model (Fig. 1c). To selectively introduce quantum corrections originating from the hydrogen nuclei while preserving the parent-PACE description of pure Fe, the Fe–Fe interactions were fixed and only interactions involving hydrogen were reoptimised. Energy differences between structures were generally excluded from training, and only a single $H_2$ free-energy anchor was included to define the energy reference. No additional DFT calculations were required for this construction. The final training dataset comprised 29,062 structures and 1,261,352 atomic environments. Local extrapolation grades[42] were evaluated, confirming that the local atomic environments generated during centroid-constrained PIMD remained within the interpolation domain of the parent PACE. This confirmed that the quantum-mean-force training labels were acquired within a configurational space reliably described by the parent model. NQC-PACE reproduced the PIMD quantum mean forces with a root-mean-square error of 3.68 meV $Å^{-1}$.

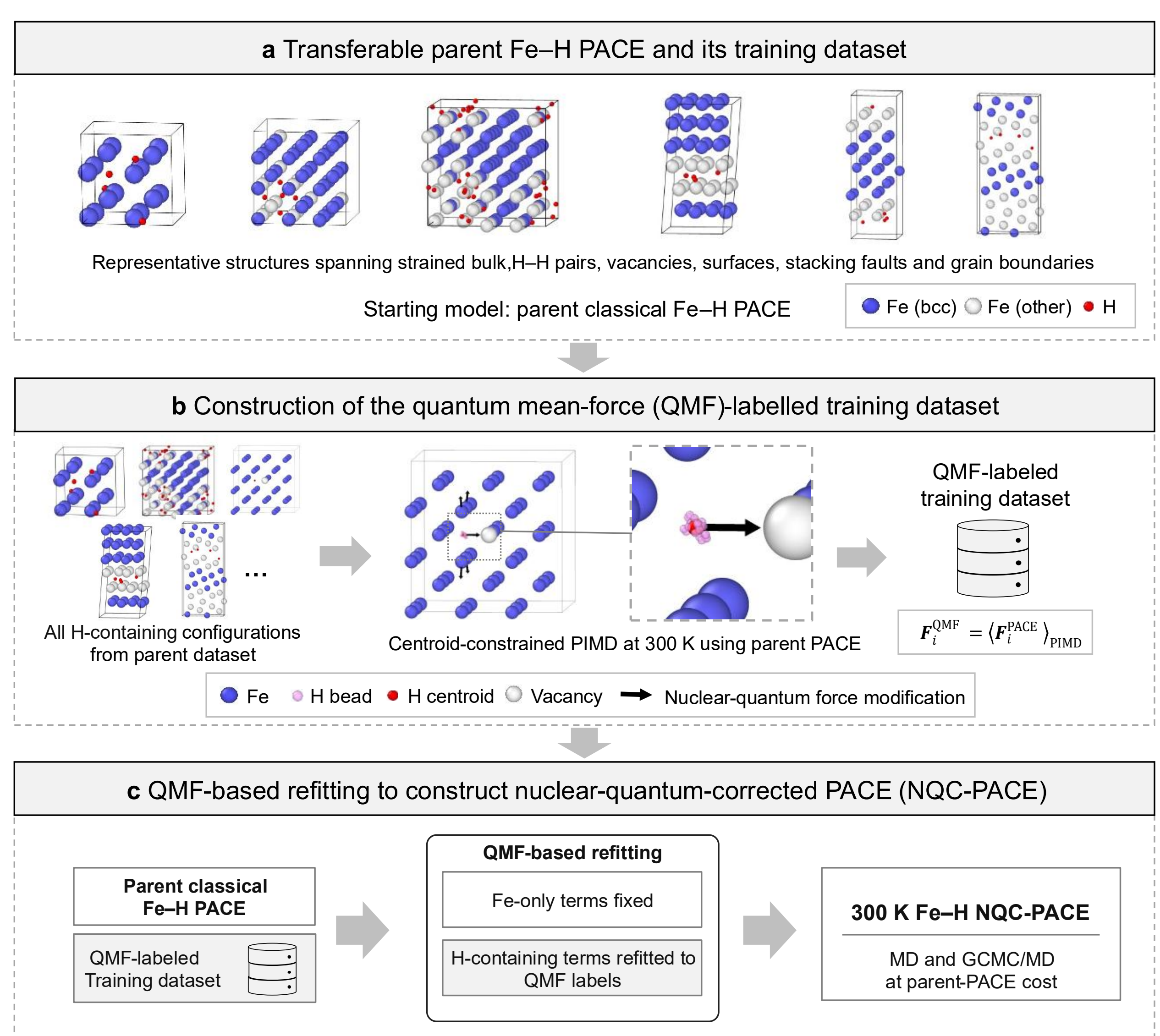

**Fig. 1 | Construction of the Fe–H nuclear-quantum-corrected PACE (NQC-PACE).**
**a**, Parent Fe–H PACE and representative training configurations spanning strained bulk, H–H pairs, vacancy–H complexes, surfaces, generalised stacking faults and grain boundaries. **b**, For each H-containing configuration, Fe atoms were fixed and the centroid of each 32-bead H ring polymer was constrained to the original H position during PIMD at 300 K. Physical forces were averaged over beads and time to generate quantum mean-force labels. Black arrows indicate the nuclear-quantum corrections to the atomic forces. **c**, Fe-only interaction terms were fixed, whereas H-containing terms were refitted to the quantum mean-force labels, yielding NQC-PACE for MD and GCMC/MD at the computational cost of the parent PACE.

NQC-PACE represents an effective energy surface that reflects quantum mean forces obtained from centroid-constrained PIMD at 300 K, in which only the hydrogen nuclei are quantised, rather than a conventional 0 K potential-energy surface. The quantum mean forces obtained from centroid-constrained PIMD correspond to the negative gradient of an effective free-energy surface at 300 K whose variables are the classically treated Fe coordinates and the hydrogen centroid coordinates[34,36,38]. Therefore, in the limit of exact reproduction of the quantum mean forces over the relevant configurational space, energy differences predicted by NQC-PACE correspond to centroid free-energy differences at 300 K. In the following, structural energy differences evaluated using NQC-PACE are interpreted as effective free-energy differences that include equilibrium nuclear quantum effects at 300 K.

**NQC-PACE reproduces quantum-corrected free-energy differences across diverse Fe–H environments and hydrogen diffusion in α-Fe**

NQC-PACE was constructed primarily using quantum mean forces as training targets and was not directly trained on the effective free-energy differences between structures examined below. We therefore first tested whether NQC-PACE captured the intended nuclear quantum corrections at 300 K by comparing its predictions with quantum free-energy differences obtained by applying mass thermodynamic integration (mass TI)[40] to the parent PACE. As an independent DFT validation, comparisons were also made with DFT plus zero-point vibrational energy (ZPE) to assess the sign and local-environment dependence of the quantum corrections. Hydrogen trapping energies were defined relative to hydrogen dissolved at a tetrahedral site in bulk α-Fe, with larger positive values indicating stronger stabilisation at the target site. Values obtained from NQC-PACE are treated below as effective trapping energies based on the hydrogen centroid free-energy surface at 300 K.

Figure 2a compares NQC-PACE with classical PACE supplemented by nuclear quantum free-energy corrections at 300 K obtained from mass TI, hereafter denoted PACE + mass TI. The test set included strained bulk α-Fe, H–H pairs in bulk, vacancies, surfaces, edge and screw dislocations, symmetric tilt grain boundaries and general grain boundaries.

The effective free-energy differences predicted by NQC-PACE closely matched the PACE + mass TI results over a broad energy range and across diverse local environments. The mean absolute error and root-mean-square error relative to perfect agreement were 2.11 and 4.20 meV, respectively. This agreement demonstrates that NQC-PACE reproduces not only the quantum mean forces used for training but also the effective free-energy differences between structures corresponding to their integrated values.

Figure 2b–e presents representative results for multiple-hydrogen trapping at a vacancy, hydrogen trapping near a screw-dislocation core, H–H interactions in α-Fe and trapping sites at a general grain boundary. The DFT results for the screw dislocation were compared with those reported previously[15]. Classical PACE accurately reproduced the changes in classical-nuclei binding energies predicted by DFT across all these environments. By contrast, NQC-PACE showed trends consistent with DFT + ZPE, in which the zero-point energies of the local hydrogen vibrational modes were included. At vacancies, dislocations and general grain boundaries, nuclear quantum effects strengthened hydrogen trapping, and NQC-PACE accordingly predicted larger trapping energies than classical PACE. For H–H interactions in bulk α-Fe, NQC-PACE reproduced the DFT + ZPE trend towards reduced effective repulsion between hydrogen atoms. Similar comparisons were performed for isolated $H_2$, surfaces, symmetric tilt grain boundaries and other validation structures, with the results presented in the Supplementary Information.

Whereas DFT + ZPE provides a zero-point vibrational correction at 0 K, NQC-PACE is an effective potential trained on PIMD mean forces at 300 K; the magnitudes of the two corrections therefore need not coincide. Indeed, for both hydrogen trapping at lattice defects and repulsive H–H interactions, the quantum corrections predicted by NQC-PACE at 300 K tended to be smaller than the corresponding 0 K corrections from DFT + ZPE. This is consistent with the decreasing difference between quantum- and classical-nuclear systems with increasing temperature. Because the comparison concerns differences in the corrections between two local environments, however, it does not imply that the NQC-PACE correction must always be smaller than the DFT + ZPE correction. Nevertheless, the two approaches agreed in the sign of quantum stabilisation or repulsion reduction and in its site dependence. These results confirm that NQC-PACE reproduces nuclear quantum corrections at 300 K not only in a specific bulk environment but also across diverse lattice defects and H–H configurations relevant to hydrogen embrittlement.

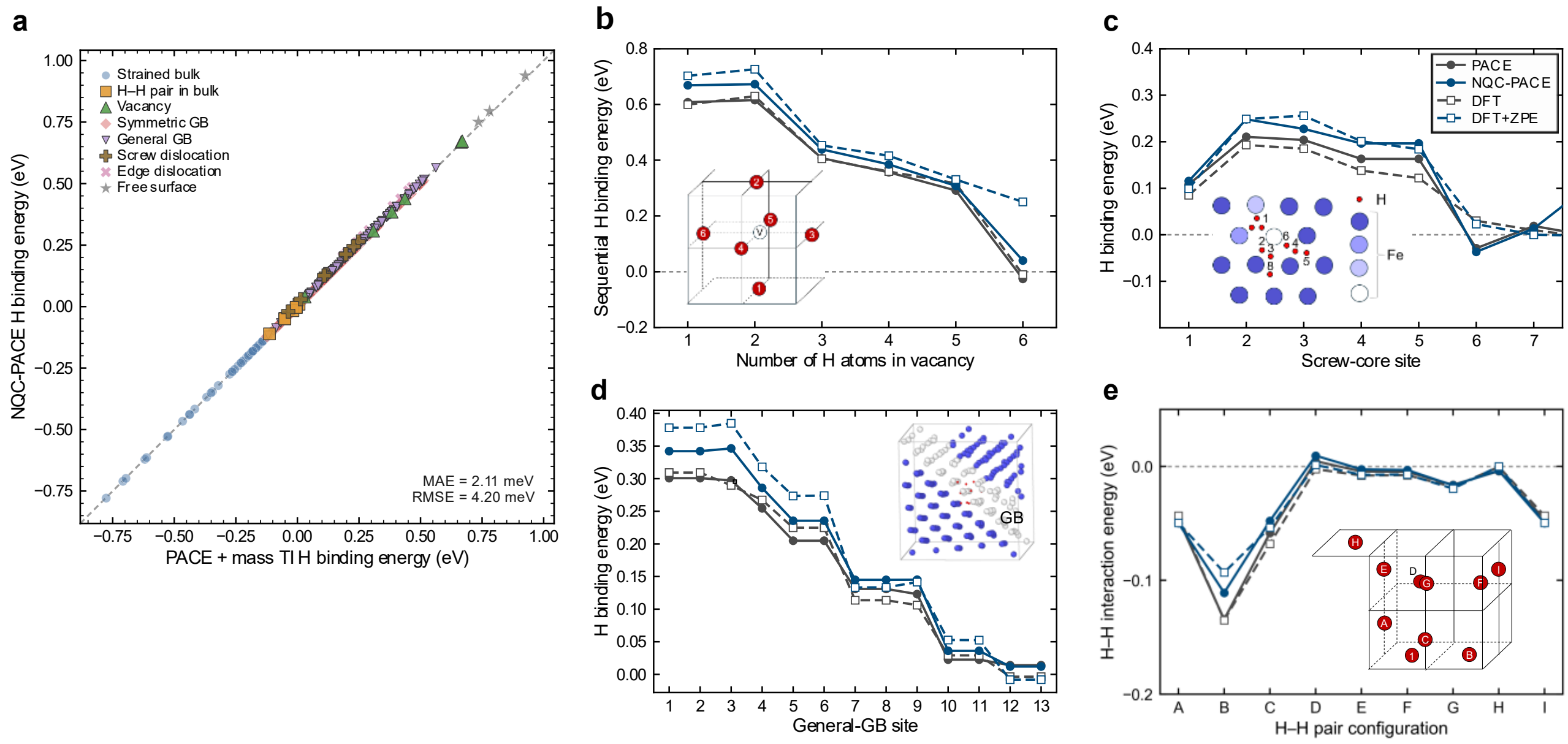

**Fig. 2 | NQC-PACE reproduces nuclear quantum effects on hydrogen trapping and H–H interactions across diverse Fe–H environments.**

**a**, Hydrogen binding energies obtained with NQC-PACE and the parent PACE combined with mass thermodynamic integration (mass TI) at 300 K for strained bulk, bulk H–H pairs, vacancies, grain boundaries, screw and edge dislocations, and free surfaces. The dashed line denotes perfect agreement; the mean absolute error (MAE) and root-mean-square error (RMSE) shown quantify deviations between the two methods over the full dataset. Their close agreement demonstrates that NQC-PACE quantitatively reproduces the 300 K nuclear quantum corrections across diverse local environments. **b–e**, Representative cases included in **a**, comparing the parent PACE, NQC-PACE, DFT and DFT with H-only harmonic zero-point-energy corrections (DFT+ZPE), the latter providing an independent 0 K reference. **b–d**, Sequential H trapping at a vacancy, sites around a screw-dislocation core and representative sites at a general grain boundary, for which both NQC-PACE and DFT+ZPE predict enhanced H trapping. For the screw-dislocation sites, the DFT values were taken from a previous DFT study[15]. **e**, Bulk H–H pair configurations, for which both approaches predict reduced H–H repulsion. For both hydrogen trapping and H–H interactions, the nuclear quantum corrections obtained with NQC-PACE at 300 K are generally smaller than the 0 K ZPE corrections, consistent with the diminishing difference between quantum and classical nuclear behaviour as temperature increases.

We next examined whether NQC-PACE could describe the effective energy surface governing hydrogen diffusion, in addition to equilibrium hydrogen trapping. Figure 3a shows the T–S–T diffusion pathway between neighbouring tetrahedral sites in α-Fe. Classical PACE, which does not include nuclear quantum effects, accurately reproduced the DFT potential-energy profile. When the free-energy profile at 300 K was evaluated using centroid-constrained PIMD with classical PACE as the underlying potential, the barrier from the tetrahedral site T to the transition state S decreased relative to the classical value and agreed with the value of approximately 0.07 eV reported in a previous path-integral calculation[32]. NQC-PACE accurately reproduced this PIMD free-energy profile.

For the pathway from a tetrahedral site T to an octahedral site O shown in Fig. 3b, NQC-PACE likewise agreed closely with the PIMD free-energy change. The same constrained pathway as that used in a previous path-integral analysis[43] was adopted for the T–O calculation. This pathway is not a minimum-energy migration path but instead probes the relative stability of the T site and an O-type environment. NQC-PACE reproduced the effective energy surface encompassing the stable T site, the O-type local environment and intermediate configurations along the pathway.

Figure 3c shows the H diffusion coefficients at 300 K obtained from conventional MD simulations. The diffusion coefficients obtained with classical PACE and NQC-PACE were $(3.11 \pm 0.05) \times 10^{-9}$ and $(5.52 \pm 0.16) \times 10^{-9}\ \mathrm{m^2\ s^{-1}}$, respectively, with NQC-PACE yielding a value approximately 1.77 times that of classical PACE. Here, the uncertainties represent the standard errors across four independent 400-ps production trajectories initiated with different velocity seeds. Whereas classical PACE yielded a diffusion coefficient lower than the experimental values, the NQC-PACE result was consistent with the range of values obtained in previous MLIP-based centroid molecular dynamics (CMD) and ring-polymer molecular dynamics (RPMD) calculations[32], and substantially improved the agreement with experiment relative to classical PACE. The experimental values shown here are those reported in previous measurements[44-49] and identified as reliable in the assessment by Kwon et al.[32]. This result does not imply that conventional MD using NQC-PACE is equivalent to exact quantum real-time dynamics, but indicates that it provides effective transport behaviour incorporating nuclear quantum effects for H in α-Fe at 300 K.

This accuracy was achieved with almost no loss of computational efficiency. As shown in Fig. 3d, obtaining a single quantum-mean-force label for a representative Fe–H configuration using PACE + PIMD required sampling for 11,000 steps with 32 beads. Based on the product of the computational resources used and wall time, the cost of obtaining one quantum-mean-force label was approximately $3.9 \times 10^5$ times that of a single NQC-PACE force evaluation. NQC-PACE and the parent PACE use the same ACE basis and functional form and therefore have the same computational cost. NQC-PACE can consequently introduce nuclear quantum effects on hydrogen diffusion, trapping at lattice defects and H–H interactions into large-scale atomistic simulations, including conventional MD and GCMC/MD, without repeating PIMD or mass TI for each analysis.

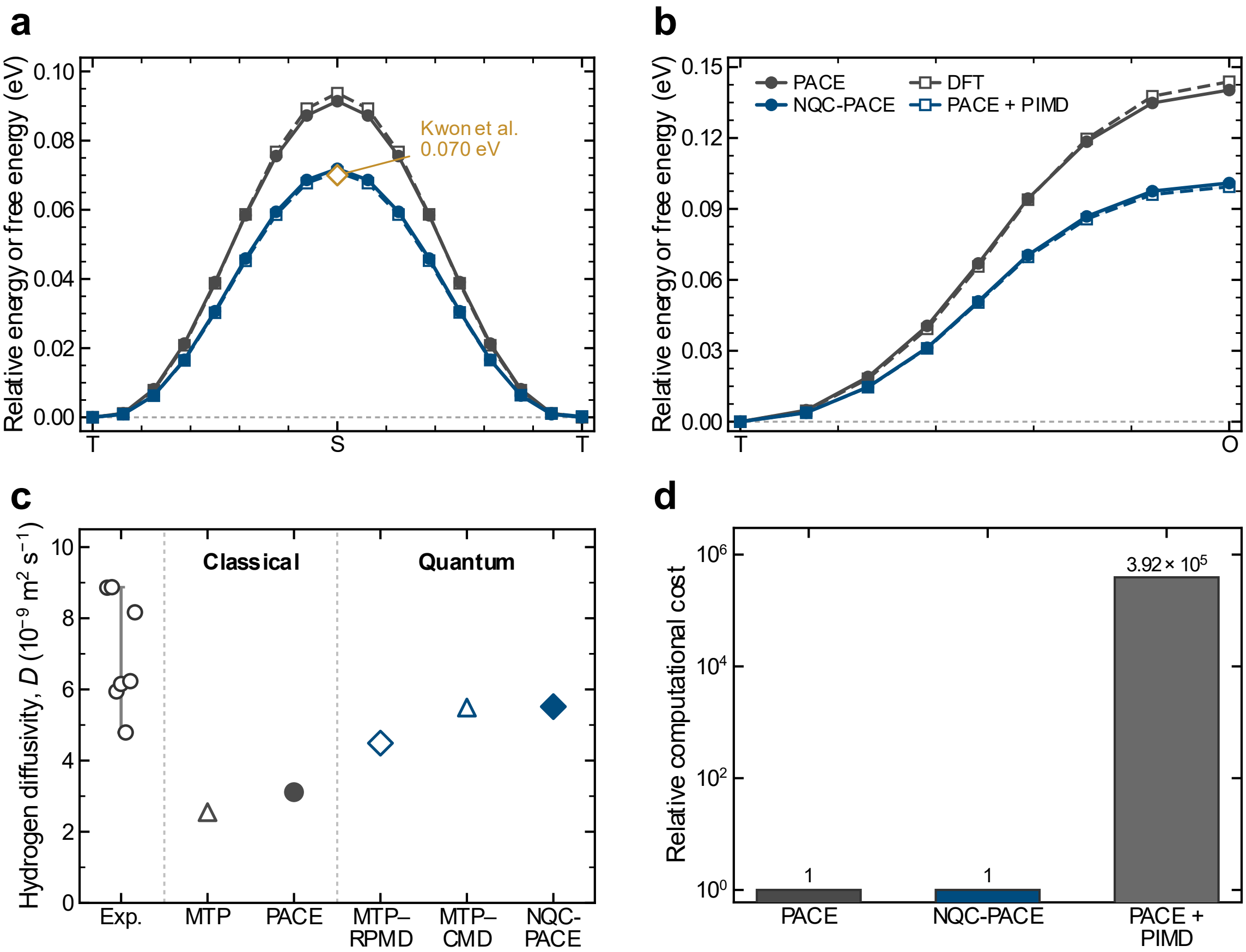

**Fig. 3 | NQC-PACE captures nuclear quantum effects on hydrogen diffusion at the same computational cost as the parent PACE.**

**a**, Relative energy and 300 K free-energy profiles along the T–S–T pathway from a tetrahedral site (T) to a neighbouring T site through the saddle point (S). Classical PACE reproduces the density functional theory (DFT) energy profile, whereas NQC-PACE agrees well with the free-energy profile obtained from parent PACE combined with path-integral molecular dynamics (PIMD) and with the previously reported path-integral result[32]. **b**, Relative energy and free-energy profiles along the constrained T–O pathway connecting the T and octahedral (O) sites. This is not a minimum-energy migration pathway but probes the relative stability of the T site and the O-like local environment. NQC-PACE reproduces the nuclear quantum stabilisation of the O-like environment obtained from PIMD. **c**, H diffusion coefficients at 300 K obtained from conventional molecular dynamics. Open circles denote reported experimental values[44-49], and the vertical line spans their range. MTP–CLMD, MTP–RPMD and MTP–CMD denote previously reported classical molecular dynamics (CLMD), ring-polymer molecular dynamics (RPMD) and centroid molecular dynamics (CMD) results obtained using a moment tensor potential (MTP), respectively[32]. NQC-PACE yields a higher diffusivity than classical PACE, consistent with experiment and previous quantum-dynamics calculations. **d**, Relative computational costs normalised to a single force evaluation using the parent PACE or NQC-PACE. Obtaining a quantum mean-force label by explicit PIMD is substantially more expensive, whereas NQC-PACE has essentially the same evaluation cost as the parent PACE.

**Enhanced hydrogen segregation at general grain boundaries due to nuclear quantum effects**

GCMC/MD simulations at 300 K were performed using NQC-PACE for six general grain boundaries in α-Fe. Figure 4a shows the grain-boundary models used in the simulations. They had misorientation angles of 25.4–47.9° and grain-boundary energies of 1.61–1.83 J $m^{-2}$, placing them within the energy distribution of high-angle grain boundaries in nanopolycrystals with random grain orientations. The mean grain-boundary energy obtained from the nanopolycrystalline model also closely agreed with the experimentally estimated energy of general grain boundaries[50]. These bicrystal models were therefore used as representative general grain boundaries in practical polycrystalline materials rather than as specific low-Σ coincidence-site-lattice boundaries.

Figure 4b shows the hydrogen-concentration profile near a representative grain boundary. For NQC-PACE, the Low, Middle and High conditions corresponded to bulk hydrogen concentrations of 0.00525, 0.0347 and 0.229 appm, respectively. For comparison, the classical-PACE result at a bulk hydrogen concentration of 0.0280 appm, which was close to that of the NQC-PACE Middle condition, is also shown. Hydrogen segregated towards the grain boundary in all simulations, but the amount of segregation at comparable bulk concentrations differed markedly depending on whether nuclear quantum effects were included. The grain-boundary hydrogen areal density averaged over the six boundaries was approximately 2.40 atom $nm^{-2}$ with NQC-PACE, compared with approximately 0.96 atom $nm^{-2}$ with classical PACE. Thus, inclusion of nuclear quantum effects increased hydrogen segregation at general grain boundaries by a factor of approximately 2.5 relative to the classical-nuclei approximation.

Figure 4c shows the relationship between the grain-boundary hydrogen areal density and the effective mean trapping energy obtained for the six general grain boundaries. The effective mean trapping energy was evaluated by time-averaging, at 300 K, the energy difference predicted by each model between the multi-hydrogen segregated state formed during GCMC/MD and the corresponding state obtained by removing all hydrogen atoms from the same Fe configuration. For NQC-PACE, this quantity is the time average of an effective energy difference on the hydrogen centroid free-energy surface at 300 K. It is not a static value for a single ideal site, but instead represents the mean stability of the segregated hydrogen population, including the inequivalent sites actually occupied, H–H interactions and finite-temperature fluctuations within those sites. Hydrogen did not migrate between trapping sites during the averaging period. Although TDS yields a single apparent trapping energy, it reflects desorption from a hydrogen population distributed over multiple grain-boundary sites and can therefore be regarded as an effective population-averaged quantity. This motivated comparison with the calculated mean stability of the segregated hydrogen population. Because thermal desorption spectroscopy (TDS) additionally includes desorption kinetics during heating, however, the calculated and experimentally derived trapping energies are not strictly identical physical quantities[7].

As the bulk hydrogen concentration increased, the grain-boundary hydrogen areal density increased, whereas the effective mean trapping energy decreased. This behaviour arises because the strongest trapping sites are occupied first, followed progressively by weaker sites. Experiments have reported a grain-boundary hydrogen areal density of 2.3–2.9 atom $nm^{-2}$ and a trapping energy of approximately 0.45 eV[7]. At a comparable hydrogen areal density, NQC-PACE predicted an effective mean trapping energy close to, although slightly higher than, the experimental value. The experimental value therefore lay between the classical-PACE and NQC-PACE predictions and was closer to NQC-PACE. By contrast, classical PACE required a higher bulk hydrogen concentration to reach the experimentally observed areal-density range and still underestimated the effective mean trapping energy under those conditions.

Figure 4d,e show the local structures of the hydrogen sites actually occupied during GCMC/MD. Based on the local Fe coordination around hydrogen, these sites were classified as tetrahedral (TET), octahedral (OCT) or pentagonal-bipyramidal (PBP) types. Hydrogen segregated predominantly to sixfold-coordinated OCT-type and fivefold-coordinated PBP-type sites with large Voronoi volumes, rather than to fourfold-coordinated TET-type sites corresponding to the stable interstitial environment in bulk α-Fe. This preferential occupation demonstrates that open, non-bulk-like local environments at general grain boundaries govern hydrogen segregation.

The nominally pure Fe used in the experiment contained 10 wt ppm carbon[7]. Even at such trace bulk concentrations, on the order of tens of weight ppm, carbon is known to segregate markedly to grain boundaries and preferentially occupy OCT-type interstitial sites[51]. As shown by the site analysis above, hydrogen also preferentially occupies OCT-type sites near general grain boundaries.

Experiments have further shown that carbon segregation suppresses hydrogen accumulation through competition for these sites[12]. Carbon in the experimental material may therefore competitively occupy some of the strongest hydrogen-trapping sites at the grain boundaries, leaving hydrogen to segregate to relatively weaker sites and thereby reducing the measured mean trapping energy. Such competition provides a plausible explanation for why the carbon-free NQC-PACE calculation slightly overestimated the experimental value. In the absence of carbon, the experimentally determined trapping energy would therefore be expected to increase towards the NQC-PACE result, while the discrepancy from classical PACE would become still larger. When this C–H site competition is considered, NQC-PACE provides a more consistent description than classical PACE of both the grain-boundary hydrogen content and the mean stability of the segregated hydrogen population.

These results demonstrate that nuclear quantum effects are important for describing both the amount of hydrogen segregated to general grain boundaries and the mean stability of the segregated hydrogen population. In particular, inclusion of nuclear quantum effects increased hydrogen segregation at general grain boundaries by approximately 2.5 times at comparable bulk hydrogen concentrations relative to the classical-nuclei approximation. We therefore next performed a comprehensive analysis of hydrogen-trapping sites at a general grain boundary to identify the structural origin of this pronounced increase in segregation.

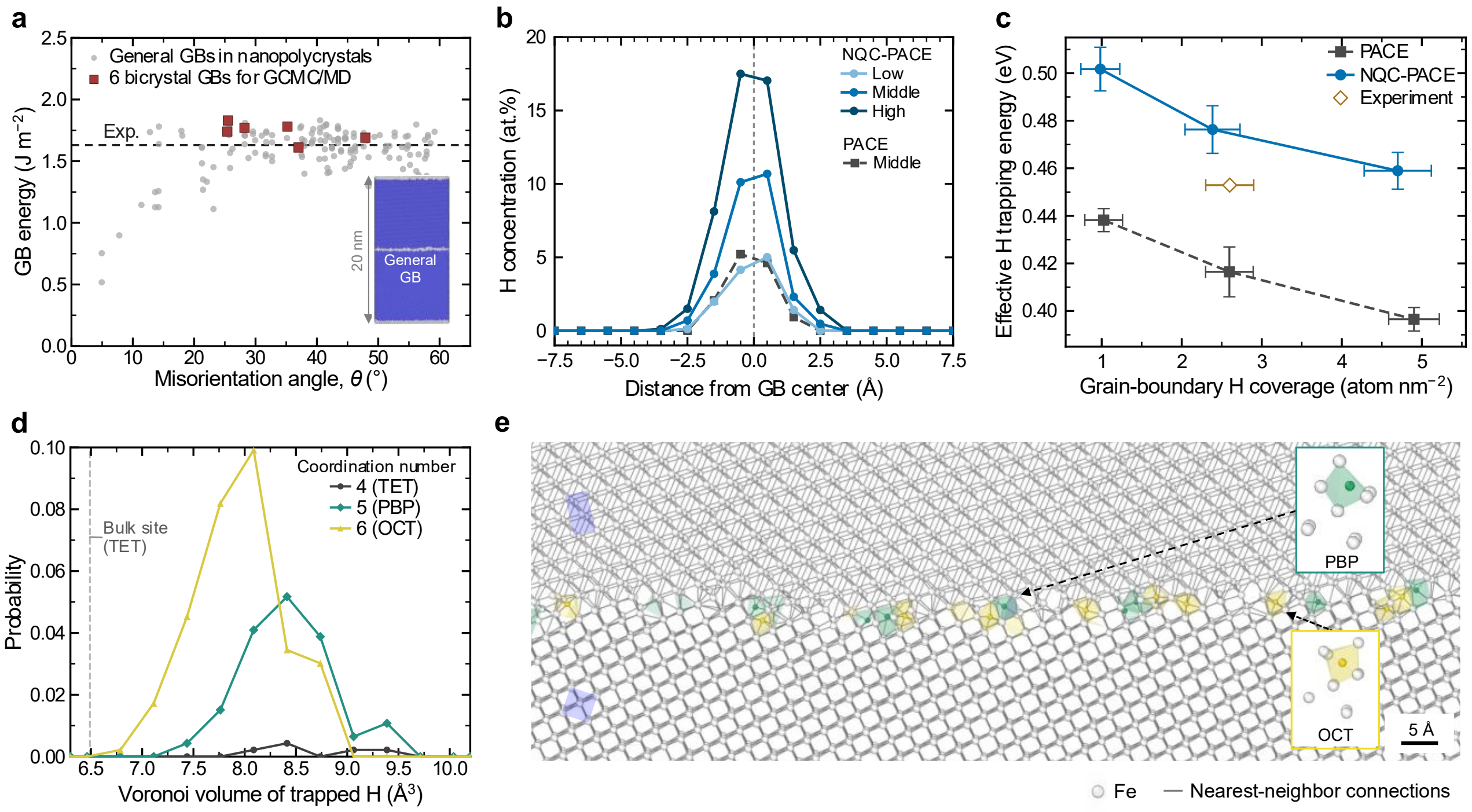

**Fig. 4 | Nuclear quantum effects enhance hydrogen segregation at general grain boundaries.** **a**, Grain-boundary (GB) energy as a function of misorientation angle for general GBs extracted from nanopolycrystalline Fe. Red squares denote the six bicrystal GBs selected for GCMC/MD, and the dashed line indicates the experimental reference. The selected GBs lie within the high-energy range characteristic of general GBs. The inset shows an example atomic structure of a selected bicrystal GB. **b**, H concentration profiles normal to the GB obtained from GCMC/MD at 300 K. For NQC-PACE, Low, Middle and High correspond to bulk H concentrations of 0.00525, 0.0347 and 0.229 appm, respectively, whereas Middle for the parent PACE corresponds to 0.0280 appm. At comparable bulk concentrations, NQC-PACE predicts markedly stronger GB segregation. **c**, Effective H trapping energy averaged over the six GBs as a function of GB H areal density. Symbols and error bars denote the mean and standard deviation across the six GBs, respectively. The experimental symbol and horizontal bar indicate the reported estimate and range[7]. Trapping weakens with increasing H density but remains stronger with NQC-PACE than with the parent PACE, with the experimental estimate lying between the two predictions. **d**, Probability distributions of the Voronoi volume around GB-segregated H, classified into fourfold tetrahedral (TET), fivefold pentagonal-bipyramidal (PBP) and sixfold octahedral (OCT) environments. **e**, Spatial distribution of the local environments occupied by segregated H in a representative general GB. Segregated H preferentially occupies the larger PBP and OCT sites formed near the boundary.

**Structural origin of site-dependent quantum stabilisation at general grain boundaries**

To determine why nuclear quantum effects enhanced hydrogen segregation at general grain boundaries, hydrogen-trapping sites near a representative boundary were analysed comprehensively. Using the interstitial-site search method for general grain boundaries described in the Methods[19], 74,554 candidate sites were generated near the boundary. A single hydrogen atom was placed at each candidate site, followed by independent relaxation using classical PACE and NQC-PACE. The quantum stabilisation was evaluated from the difference in trapping energy between the two models for the same initial candidate site.

Figure 5a shows the relationship between the hydrogen trapping energy predicted by NQC-PACE and the hydrogen Voronoi volume. The trapping energy tended to increase with local volume, with particularly strong hydrogen trapping obtained at OCT- and PBP-type sites having large Voronoi volumes. These regions also corresponded to the distribution of sites actually occupied by hydrogen during the GCMC/MD simulations in Fig. 4.

Figure 5b shows the trapping-energy difference between NQC-PACE and classical PACE for the same site,

$$\Delta E_{\mathrm{Q}} = E_{\mathrm{trap}}^{\mathrm{NQC}} - E_{\mathrm{trap}}^{\mathrm{PACE}}.$$

Here, $\Delta E_{\mathrm{Q}}$represents the site-dependent quantum stabilisation on the hydrogen centroid free-energy surface at 300 K relative to the classical-PACE potential-energy surface. $\Delta E_{\mathrm{Q}}$was positive for most sites, indicating that nuclear quantum effects strengthened hydrogen trapping at the grain boundary.

Its magnitude also tended to increase with Voronoi volume. However, even when sites with comparable Voronoi volumes were compared, $\Delta E_{\mathrm{Q}}$ was larger for OCT-type sites than for TET-type sites. The magnitude of quantum stabilisation is therefore governed not only by the local free volume but also by the coordination geometry of the surrounding Fe atoms.

To examine this coordination dependence, Fig. 5c shows the local structures and hydrogen-displacement potential-energy surfaces of a grain-boundary TET-type site and an OCT-type site with similar Voronoi volumes, together with an OCT-type site having a still larger volume. These potential-energy surfaces were evaluated using classical PACE. The TET-type site exhibited a relatively isotropic potential well, whereas the OCT-type site formed an anisotropic well with low curvature along a particular direction. At the large-volume OCT-type site, the potential was softened along multiple directions.

Figure 5d shows the quantum free-energy correction at 300 K for each vibrational mode, evaluated within the harmonic approximation from the curvature along the principal axes of the potential-energy surface. Along the low-curvature direction of an OCT-type site, the hydrogen vibrational frequency decreased, and the quantum free-energy correction associated with that mode became smaller than that at the bulk TET site. Relative to the bulk TET reference, this difference in the correction acted to increase the trapping energy of the OCT-type site. At the large-volume OCT-type site, several vibrational modes were softened, producing a still larger stabilisation. The site-to-site quantum free-energy differences obtained from the harmonic analysis were consistent with the trend in $\Delta E_{\mathrm{Q}}$ derived from the difference between NQC-PACE and classical PACE. For representative sites, the NQC-PACE $\Delta E_{\mathrm{Q}}$ and the harmonic corrections were also quantitatively similar.

These results show that nuclear quantum stabilisation at general grain boundaries does not act uniformly across all sites, but instead becomes enhanced in OCT- and PBP-type environments that possess both a large local free volume and anisotropically soft confinement. This site-dependent quantum stabilisation provides a structural explanation for the pronounced increase in hydrogen segregation at general grain boundaries relative to the classical-nuclei approximation.

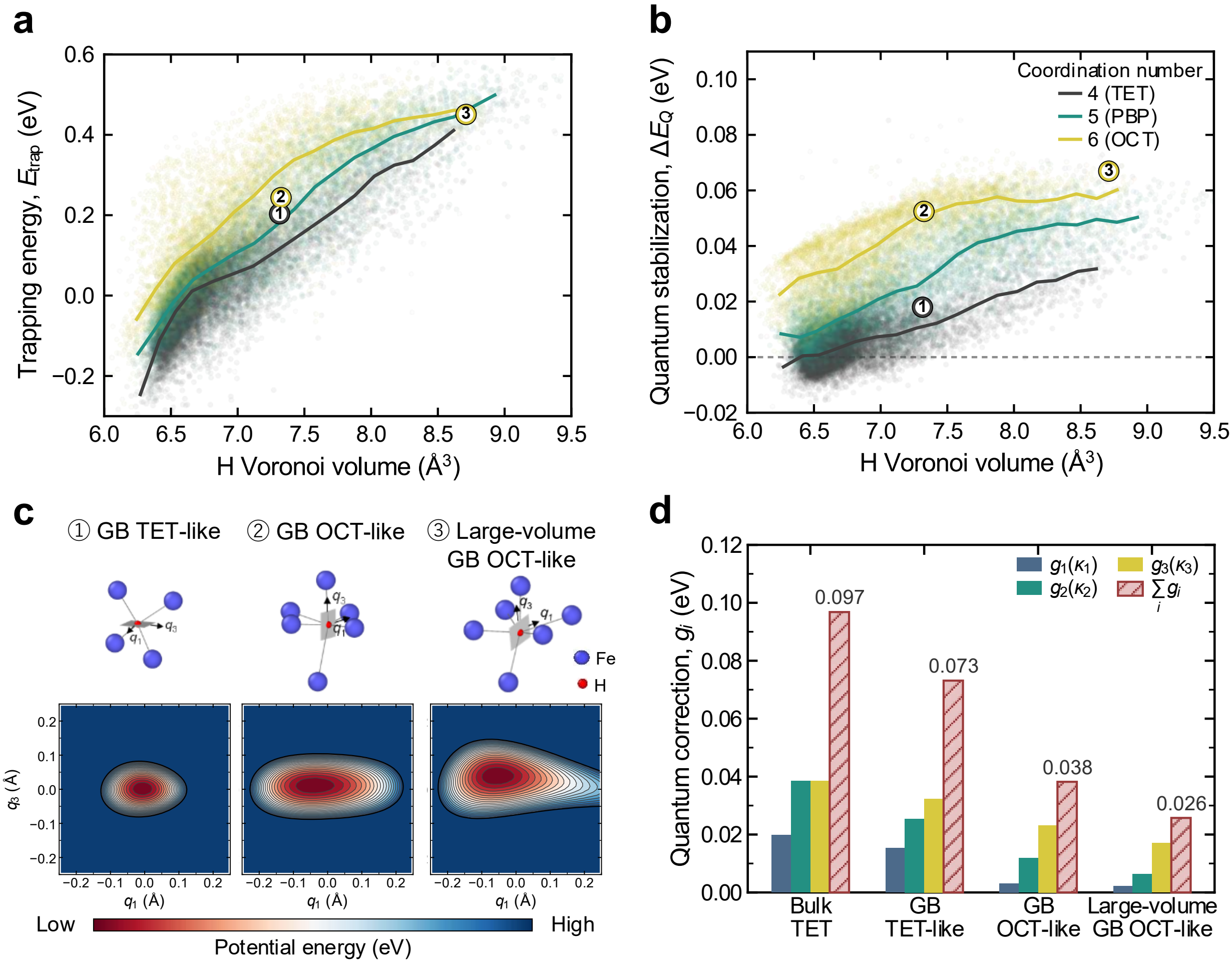

**Fig. 5 | Locally expanded OCT-like grain-boundary environments induce anisotropic H vibrational softening and enhance nuclear quantum stabilisation.**

**a, b,** H trapping energy, $E_{\mathrm{trap}}$ (**a**), and nuclear quantum stabilisation, $\Delta E_Q = E_{\mathrm{trap}}^{\mathrm{NQC\text{-}PACE}} - E_{\mathrm{trap}}^{\mathrm{PACE}}$ (**b**), as functions of the H Voronoi volume for candidate sites at a representative general grain boundary. Points denote individual sites, and solid lines show binned medians for fourfold tetrahedral (TET), fivefold pentagonal-bipyramidal (PBP) and sixfold octahedral (OCT) environments. Strong trapping and large positive $\Delta E_Q$ preferentially occur in locally expanded OCT-like environments. Numbered markers correspond to the representative sites in **c**.

**c,** Local atomic structures and parent-PACE potential-energy surfaces projected onto the $q_1$–$q_3$ mode plane for a GB TET-like site, a GB OCT-like site of comparable Voronoi volume and a large-volume GB OCT-like site. Compared with the relatively isotropic TET-like site, the OCT-like environments exhibit pronounced anisotropic softening of H confinement, which is further enhanced by local volume expansion.

**d,** Mode-resolved harmonic nuclear quantum corrections, $g_i(\kappa_i)$, at 300 K and their sums for the bulk TET site and representative GB environments. The total local correction decreases from 0.097 eV for the bulk TET site to 0.026 eV for the large-volume GB OCT-like site. This reduction relative to the bulk reference produces a positive $\Delta E_Q$ and strengthens GB trapping.

## Discussion

We constructed NQC-PACE by reusing an existing Fe–H PACE capable of describing broad atomic environments relevant to hydrogen embrittlement, together with the training configurations underpinning its structural coverage, and relabelling those configurations with quantum mean forces from PIMD at 300 K. Rather than repeating explicit PIMD for each target system, the method maps equilibrium quantum mean forces from centroid-constrained PIMD onto a temperature-dependent effective potential defined in the original atomic representation. This preserves the configurational coverage and efficiency of the parent PACE while incorporating nuclear quantum corrections into hydrogen trapping, H–H interactions and diffusion. Learning quantum centroid forces, their differences from classical forces, or PIMD-derived corrections to existing potentials has previously been proposed mainly for molecular systems[33-38]. The distinctive feature here is the direct relabelling of training structures from an existing materials MLIP spanning diverse defects and interfaces, enabling hydrogen relative stability, diffusion and segregation to be treated within one model. Reusing both the parent potential and its training configurations eliminates additional DFT calculations. Each quantum-mean-force label also requires only one centroid-constrained PIMD calculation, reducing data-generation costs relative to applying mass TI[40], which requires calculations at multiple fictitious masses, to every training configuration.

NQC-PACE does not reproduce a conventional 0 K potential-energy surface, but an effective energy surface associated with equilibrium quantum mean forces at 300 K. This interpretation is supported by the close agreement with quantum free-energy differences obtained from the parent PACE combined with mass TI. Despite differences in temperature and approximation, comparison with DFT plus zero-point vibrational energy (ZPE) also yielded consistent signs and local-environment dependence of the corrections for vacancies, dislocations, general grain boundaries and H–H interactions. For both defect trapping and repulsive H–H interactions, the 300 K corrections predicted by NQC-PACE tended to be smaller than the corresponding 0 K DFT + ZPE corrections, consistent with the diminishing difference between quantum- and classical-nuclear systems with increasing temperature[29,30]. Thus, training primarily on forces can reproduce their integrated effective free-energy differences across diverse Fe–H environments.

NQC-PACE reproduced hydrogen-diffusion free-energy profiles obtained from centroid-constrained PIMD, while conventional MD yielded a diffusion coefficient at 300 K consistent with previous CMD and RPMD calculations[32] and experiment. Classical MD with NQC-PACE is not formally equivalent to approximate quantum-dynamical methods such as RPMD[52,53]; it instead describes classical evolution on a quantum-corrected effective energy surface. Nevertheless, agreement in both the free-energy profile and diffusion coefficient supports its use as an effective transport approximation for hydrogen in α-Fe at 300 K.

In GCMC/MD simulations of general grain boundaries, NQC-PACE increased the grain-boundary hydrogen areal density by approximately 2.5 times relative to classical PACE at comparable bulk concentrations. Classical-nuclei simulations may therefore substantially underestimate grain-boundary hydrogen accumulation, particularly under dilute conditions. Even modest site-specific free-energy corrections can strongly alter occupancies in this regime and become amplified into large differences in macroscopic segregation.

The experimentally reported trapping energy lay between the NQC-PACE and classical-PACE predictions. The experimental material contained trace carbon[7], and both carbon and hydrogen preferentially occupy OCT-type sites near grain boundaries[28]. Atomistic and thermodynamic studies indicate that they may compete for the same or neighbouring strong segregation sites[54,55], while experiments show that grain-boundary carbon segregation suppresses hydrogen accumulation through site competition[12]. Carbon may therefore occupy some of the strongest hydrogen traps, leaving hydrogen to populate weaker sites and reducing the experimentally inferred mean trapping energy. In carbon-free material, the measured trapping energy could consequently be higher and closer to the NQC-PACE prediction. Although the calculated effective mean trapping energy and the value derived from thermal desorption spectroscopy are not strictly identical physical quantities, accounting for C–H competition makes NQC-PACE more consistent with the experimental trend than classical PACE.

Site-resolved analysis showed that nuclear quantum effects do not stabilise all grain-boundary sites uniformly, but are enhanced in environments with large local free volume and OCT- or pentagonal-bipyramidal-type coordination. Differences among sites with similar Voronoi volumes show that free volume alone is insufficient to explain the magnitude of stabilisation. At OCT-type sites, the local potential for hydrogen becomes anisotropically softer and develops low-frequency modes. Relative to the bulk TET site, this mode softening reduces the quantum free-energy correction and thereby strengthens trapping in the non-bulk environment. Because enhanced trapping was also obtained at vacancies and dislocations, similar site-dependent mechanisms may operate at defects other than general grain boundaries. NQC-PACE therefore provides a practical route for surveying defect-specific trapping landscapes and identifying the structural factors governing quantum stabilisation.

The applicability of this framework depends on whether the underlying MLIP and training dataset adequately cover the local environments of interest. Here, we reused structures sampled by concurrent learning across environments relevant to hydrogen embrittlement and confirmed that configurations generated during PIMD remained within the interpolation domain of the parent PACE. Reproduction of quantum free-energy differences for dislocations and general grain boundaries not explicitly included in the training dataset further supports this design. The present model is specific to 300 K, and transferability to other temperatures or environments outside the parent-PACE domain is not guaranteed. Models for other temperatures could, however, be constructed by generating quantum mean forces at the target temperature. Only hydrogen nuclei were quantised, whereas Fe nuclei were treated classically. Fe zero-point motion and quantum delocalisation are therefore neglected, but their contributions to hydrogen trapping and diffusion at 300 K are expected to be secondary because Fe is much heavier than hydrogen.

Finally, NQC-PACE approximates the mean-force surface obtained from centroid-constrained PIMD and does not rigorously reproduce general quantum nuclear position distributions or exact real-time quantum dynamics. Nevertheless, given a high-quality MLIP and training dataset with adequate configurational coverage, relabelling the same space with finite-temperature quantum mean forces should be transferable beyond Fe–H to other materials in which light-element nuclear quantum effects are important. By distilling explicit quantum-nuclear simulations into a conventional MLIP, this framework enables finite-temperature nuclear quantum effects across defects, interfaces and diffusion processes to be examined at previously inaccessible spatial and temporal scales.

## Methods

### DFT calculations and atomistic simulations

DFT calculations were performed using the Vienna *Ab initio* Simulation Package (VASP) [56,57] with the Perdew–Burke–Ernzerhof generalised-gradient approximation[58]. The plane-wave cut-off energy, *k*-point sampling, and convergence criteria for the electronic and ionic degrees of freedom were identical to those used to construct the previously reported Fe–H PACE models[27,28].

Structural relaxations, nudged elastic band (NEB) calculations, molecular dynamics (MD), path-integral molecular dynamics (PIMD), mass thermodynamic integration (mass TI), and grand-canonical Monte Carlo/molecular dynamics (GCMC/MD) simulations were performed using LAMMPS[59]. The parent potential for NQC-PACE was the previously developed transferable Fe–H PACE[27]. This potential has been validated for diverse Fe–H environments, including bulk and strained structures, H–H interactions, vacancies, dislocations, surfaces, symmetric tilt grain boundaries and general grain boundaries[27]. Details of its construction and accuracy against DFT are reported elsewhere[27]. Atomic configurations were visualised and analysed, including by Voronoi analysis, using OVITO[60].

### PIMD and quantum mean forces

PIMD calculations used the parent Fe–H PACE as the underlying potential. Only hydrogen atoms were treated as quantum nuclei, with each hydrogen nucleus represented by a ring polymer containing 32 beads. A previous ring-polymer molecular dynamics study of hydrogen diffusion and vacancy trapping in α-Fe used 16 beads at 300 K and increased the number to 64 as the temperature was lowered. To provide a conservative imaginary-time discretisation across the broader range of Fe–H environments examined here, we therefore employed 32 beads at 300 K[30]. Fe nuclei were not quantised. The internal normal modes of the hydrogen ring polymers were thermostatted at 300 K using PILE-L[31]. The equations of motion were integrated using the BAOAB Langevin splitting scheme with a time step of 1 fs. To avoid introducing arbitrary initial bead displacements that could transiently sample extrapolative short-range Fe–H configurations, all beads were initially placed at the corresponding centroid position, giving a collapsed-ring configuration. The internal ring-polymer modes were subsequently thermalised using PILE-L during the equilibration period, from which no training labels were collected. Analysis-specific conditions, including centroid constraints and changes in the hydrogen mass, are described below.

To generate training labels for NQC-PACE, Fe atoms were fixed at their positions in each parent-PACE training configuration, and the centroid of each hydrogen ring polymer was constrained to the corresponding hydrogen position[35,36]. The quantum mean force acting on atom $i$, $F_i^{\mathrm{Q}}$, was obtained by averaging the physical interatomic force over the beads and the PIMD trajectory:

$$F_i^{\mathrm{Q}} = \left\langle \frac{1}{P} \sum_{s=1}^{P} F_i^{(s)} \right\rangle,$$

where $P = 32$and $F_i^{(s)}$is the physical force acting on atom $i$for the configuration of bead $s$. Ring-polymer spring forces and forces arising from the centroid constraints were excluded. In addition to the mean forces acting on hydrogen, the mean forces induced on the surrounding Fe atoms by hydrogen quantum fluctuations were included as training targets.

For each training configuration, PIMD was run for 16,000 steps. The first 6,000 steps were discarded for equilibration, and the physical forces were averaged over the beads and the remaining 10,000 steps to obtain the quantum mean forces. The production interval was divided into twenty blocks of 500 steps, and the standard errors of the hydrogen mean forces were evaluated from the variation among the block averages.

For the 29,018 structures used to fit the quantum mean forces, the maximum local extrapolation grade over all beads and sampled frames in the production interval was below 2. The standard error of the hydrogen mean forces was below 0.02 eV $\text{Å}^{-1}$ for every structure, with a 95th-percentile value of 0.00416 eV $\text{Å}^{-1}$. These results confirmed that the quantum-mean-force labels were obtained within the interpolation domain of the parent PACE and that their statistical uncertainties were sufficiently controlled.

**Construction of NQC-PACE**

NQC-PACE was constructed by reusing the atomic configurations in the parent-PACE training dataset and relabelling them with quantum mean forces obtained from centroid-constrained PIMD at 300 K. No additional DFT calculations were performed, thereby introducing finite-temperature nuclear quantum corrections within the Fe–H configurational space described by the parent model.

To preserve the parent-PACE description of pure Fe, interaction terms involving only Fe were fixed, and only terms involving hydrogen were reoptimised. Quantum mean forces were used as the primary training targets, whereas energies and stresses were otherwise assigned zero weight. Because force-only fitting does not determine the additive constant of the free energy, one isolated $H_2$ configuration at $r_{HH} = 0.750$ Å, close to its equilibrium bond length, was assigned the 300 K free energy obtained from mass TI as an energy anchor. All other structures had zero energy weight. This single anchor fixed the energy offset without altering the shape of the $H_2$ free-energy curve determined by the quantum mean forces.

The final dataset comprised 29,062 structures and 1,261,352 atomic environments. Fitting was performed using Pacemaker[61]. NQC-PACE used the same ACE basis as the parent model, with cut-off radii of 6.0 Å for Fe–Fe interactions and 4.5 Å for interactions involving hydrogen. The final model contained 544 ACE basis functions[27]. Local extrapolation grades[42] were evaluated for environments generated during PIMD, confirming that the configurations used for training remained within the interpolation domain of the parent PACE. NQC-PACE reproduced the PIMD quantum mean forces with a root-mean-square error of 3.68 meV $Å^{-1}$.

NQC-PACE was used as an effective potential reproducing PIMD quantum mean forces at 300 K rather than as a conventional 0 K Born–Oppenheimer potential. Energy differences obtained from NQC-PACE were therefore interpreted as quantum-corrected effective relative stabilities at 300 K.

**Validation of quantum-corrected hydrogen stability**

To test whether NQC-PACE reproduced quantum-corrected energy differences involving hydrogen, we examined strained bulk α-Fe, H–H pairs in bulk, vacancies, symmetric tilt and general grain boundaries, screw and edge dislocations, and surfaces. For each environment, NQC-PACE predictions were compared with those obtained using the parent PACE combined with mass TI and DFT plus zero-point vibrational energy (ZPE).

Hydrogen trapping energies were defined as the difference between the hydrogen insertion energies at a bulk tetrahedral site and at the target defect site, such that a positive value denotes defect-induced stabilisation. H–H interaction energies were evaluated relative to two dissolved hydrogen atoms separated sufficiently far to be non-interacting.

For mass TI[40], the PIMD settings described above were used while progressively increasing the hydrogen nuclear mass from its physical value towards the classical limit. The free-energy difference between the quantum- and classical-nuclear systems was obtained by integrating the centroid-virial kinetic energy[40] over mass. Thirteen mass points spanning mass factors from 1 to 64 were used, and Simpson's rule was applied for numerical integration. Sensitivity to the integration procedure, statistical uncertainty and intratrajectory drift was assessed, confirming that these conditions were sufficient for the present free-energy differences. PACE + mass TI trapping energies were obtained by adding the difference between the quantum free-energy corrections at the bulk and defect sites to the classical-PACE trapping energy.

DFT + ZPE validation used atomic structures previously relaxed and evaluated by DFT in the Fe–H neural-network-potential study28; the DFT settings and relaxation procedures are reported there. New finite-displacement harmonic vibrational analyses were performed for the hydrogen degrees of freedom only. DFT + ZPE is a 0 K harmonic approximation and is not identical to the 300 K correction represented by the PIMD mean forces in NQC-PACE. It was therefore used as an independent test of the sign and relative environmental dependence of the quantum corrections.

To reduce computational cost while retaining the dominant hydrogen contribution, Fe atoms were fixed and only hydrogen vibrational degrees of freedom were included. This was particularly important for the general-grain-boundary models containing approximately 250 Fe atoms[28]. For structures containing multiple hydrogen atoms, all hydrogen degrees of freedom were retained so that vibrational coupling between hydrogen atoms was included. ZPE values were calculated from the resulting local hydrogen modes, and the corresponding ZPE difference between the defect and bulk tetrahedral environments was incorporated into the DFT trapping energy.

**Validation against hydrogen diffusion**

Hydrogen diffusion in bcc Fe was examined along the T–S–T pathway from one tetrahedral site to a neighbouring tetrahedral site through the saddle point, and along the T–O pathway from a tetrahedral to an octahedral site. The T–S–T pathway was obtained with classical PACE using the NEB method[62], whereas the T–O pathway was constructed by constrained relaxation.

Diffusion energy and free-energy profiles were evaluated following a previous path-integral study[43].

For the T–O pathway, hydrogen was constrained along the reaction coordinate joining the T and O sites, as in that study. This is not a minimum-energy migration pathway, but a constrained path probing the relative stability of the T and O environments, which are differently affected by quantum fluctuations.

At each reaction coordinate, centroid-constrained PIMD was performed and the quantum mean force along the coordinate was numerically integrated to obtain the free-energy profile at 300 K. For NQC-PACE, both effective energy differences and profiles obtained by integrating forces along the same paths were evaluated and compared with the free-energy profiles obtained from PIMD driven by the parent PACE.

Diffusion coefficients at 300 K were calculated for comparison with the previous hydrogen-diffusion study[32] using a periodic $Fe_{8192}H_{64}$ supercell with the same H/Fe ratio of 1/128. NVT-MD simulations were performed at the respective 300 K volumes of classical PACE and NQC-PACE, with a Nosé–Hoover thermostat applied to all Fe and H atoms. After equilibration for 100 ps, four independent 400-ps production trajectories were generated using different initial velocities. The diffusion coefficients were calculated from the mean-squared displacement of H using the Einstein relation, and the uncertainties were evaluated as the standard errors over the four independent trajectories.

The computational costs of explicit PIMD and NQC-PACE were compared using an $Fe_{54}H_1$configuration near the T–S–T transition state. For PACE + PIMD, one quantum-mean-force label was obtained from 11,000 steps with 32 beads; for NQC-PACE, a single force evaluation was performed for the same configuration. Computational cost was compared using the product of the computational resources used and wall-clock time.

**Hydrogen segregation at general grain boundaries**

Hydrogen segregation was analysed using six α-Fe general-grain-boundary bicrystal models reported previously[28]. Each model had approximate dimensions of 10 nm × 10 nm × 20 nm, with periodic boundary conditions parallel to the grain boundary and a non-periodic direction normal to it. Details of model construction and structural relaxation were reported previously[28]. Before the grain-boundary simulations, bulk α-Fe GCMC/MD calculations were performed independently for classical PACE and NQC-PACE to calibrate the relationship between hydrogen chemical potential and equilibrium bulk concentration. Because the two potentials have different energy references, results were compared at corresponding bulk hydrogen concentrations rather than at identical chemical potentials.

GCMC/MD simulations of general grain boundaries were performed in the NVT ensemble at 300 K. Hydrogen insertion and deletion were restricted to within 1 nm of the grain-boundary centre. Every 100 MD steps, 500 hydrogen insertion or deletion attempts were performed. Hydrogen atoms were unconstrained during the subsequent MD and could migrate freely near the boundary.
Each condition was simulated for 500,000 steps. Equilibration was assessed from the time evolution of the hydrogen population and grain-boundary hydrogen areal density, and equilibrated snapshots and time-averaged quantities were analysed. The areal density was calculated by dividing the number of segregated hydrogen atoms by the grain-boundary area. The standard deviation among the six boundaries was used to quantify variations arising from grain-boundary structure.

For comparison with trapping energies derived from thermal desorption spectroscopy (TDS) using the Choo–Lee analysis[63], an effective mean trapping energy per hydrogen atom was evaluated for the multi-hydrogen segregation states generated by GCMC/MD. Because the experimental value reflects desorption from hydrogen distributed among multiple grain-boundary sites, it does not necessarily correspond to the static trapping energy of an individual ideal site. We therefore used the mean stability of the segregated state, including inequivalent site occupation and H–H interactions, as the comparison metric.

For each equilibrated hydrogen-containing grain-boundary snapshot, a corresponding configuration was generated by removing all hydrogen atoms from the same Fe structure. Independent MD simulations at 300 K were then performed for the hydrogen-containing and hydrogen-free configurations using a time step of 0.5 fs for 10 ps, with the first 1 ps discarded. For a representative boundary, cumulative averages were compared while extending the analysis endpoint from 2 to 10 ps. For both classical PACE and NQC-PACE, changes after 3 ps were within 0.006 eV per H; 1–10 ps was therefore used as the standard averaging window. Hydrogen did not migrate between trapping sites during this averaging period.

The effective mean trapping energy per hydrogen atom was defined as

$$E_{\mathrm{trap}}^{\mathrm{eff}} = E_{\mathrm{ins}}^{\mathrm{bulk}} - \frac{\langle E_{\mathrm{GB+H}} \rangle - \langle E_{\mathrm{GB}} \rangle}{N_{\mathrm{H}}},$$

where $E_{\mathrm{ins}}^{\mathrm{bulk}}$ is the hydrogen insertion energy at a bulk tetrahedral site and $N_{\mathrm{H}}$ is the number of hydrogen atoms at the boundary. This is not a static single-site trapping energy, but a time-averaged quantity for a multi-hydrogen segregation state that includes occupation of inequivalent sites, H–H interactions, grain-boundary relaxation and finite-temperature fluctuations within the occupied sites at 300 K.

**Analysis of site-dependent quantum stabilisation**

For site-resolved analysis, a grain boundary showing segregation behaviour close to the average of the six boundaries was selected. All hydrogen atoms were removed from an equilibrated NQC-PACE GCMC/MD snapshot, after which the Fe coordinates were relaxed using NQC-PACE while retaining the cell dimensions corresponding to 300 K. This relaxed structure was used for candidate-site generation.

Within ±5 Å normal to the grain-boundary plane, vertices of Voronoi polyhedra constructed from Fe atoms were generated as tetrahedral-like candidates, and the midpoints of second-nearest-neighbour Fe–Fe pairs were generated as octahedral-like candidates, following the previously developed interstitial-site search method for general grain boundaries[19]. After merging nearby candidates, 74,554 sites remained.

One hydrogen atom was placed at each candidate site, followed by independent structural relaxation with classical PACE and NQC-PACE. Trapping energies were evaluated for each potential relative to hydrogen dissolved at a bulk tetrahedral site. The additional trapping stabilisation due to nuclear quantum effects was defined as

$$\Delta E_{\mathrm{Q}} = E_{\mathrm{trap}}^{\mathrm{NQC}} - E_{\mathrm{trap}}^{\mathrm{PACE}}.$$

Voronoi analysis of the relaxed hydrogen sites was performed using OVITO. Sites with hydrogen Voronoi coordination numbers of 4, 5 and 6 were classified as tetrahedral (TET), pentagonal-bipyramidal (PBP) and octahedral (OCT) types, respectively.

For representative sites, the Fe configuration relaxed with classical PACE was fixed and hydrogen was displaced slightly around its equilibrium position to construct the local Hessian for the three hydrogen degrees of freedom. The principal directions and local vibrational frequencies, $\omega_k$, were obtained from the eigenvalues of the mass-weighted Hessian. For each mode, the free-energy difference between quantum and classical harmonic oscillators at 300 K was evaluated as

$$\Delta F_k^{\mathrm{Q-cl}}(T) = k_{\mathrm{B}} T \ln \left[ \frac{2 \sinh \ (\hbar\omega_k / 2 k_{\mathrm{B}} T)}{\hbar\omega_k / k_{\mathrm{B}} T} \right].$$

The difference in these corrections between a grain-boundary site and the bulk TET site was compared with $\Delta E_{\mathrm{Q}}$obtained from NQC-PACE and classical PACE to identify the structural origin of site-dependent quantum stabilisation.

**Data availability**

The final NQC-PACE interatomic potential developed in this study is publicly available at https://github.com/KazumaIto0810/Fe-H_NQC-PACE. The training dataset used to construct the potential is not publicly available because of intellectual property restrictions. Source data for the figures and representative atomic structures are available from the corresponding author upon reasonable request.

**Code availability**

The simulations and potential training used publicly available software packages, including LAMMPS[59] and Pacemaker[61].

**Acknowledgments**

This work used computational resources of the Supercomputer Fugaku provided by Riken through the HPCI System Research Project (Project IDs: hp230272, hp240280, hp250310). This work was partly supported by Accompanying User Support Program (【23Z-03, 23Z-05, 24H1-01】, Support content:【porting of application program, execution performance tuning】) performed by Research Organization for Information Science and Technology.

**Funding statement**

This research received no external funding.

**Author contributions**

K.I. conceived the study, developed and validated the Fe–H MLIP, performed the simulations, analysed the data, prepared the figures and wrote the manuscript.

**Competing interests**

The author declares no competing interests.